\documentclass[letterpaper]{article}
\usepackage{aaai}
\usepackage{times}
\usepackage{helvet}
\usepackage{courier}
\usepackage{graphicx}
\usepackage{amsmath}
\usepackage{amssymb}
\usepackage{multirow}
\usepackage{xcolor}
\usepackage{float}
\usepackage{todonotes}
\usepackage{comment}
\usepackage{xspace}
\usepackage[normalem]{ulem}

\newcommand{\eg}{\textit{e}.\textit{g}. }
\newcommand{\etc}{\textit{etc}. }

\newcommand{\me}[1]{#1\xspace}
\newcommand{\says}[3]{}

\begin{document}
%
\title{\textbf{NeuronInspect}: Detecting Backdoors in Neural Networks via Output Explanations}
\author{Xijie Huang$^{1}$\thanks{Xijie Huang is from Shanghai Jiao Tong University. This work is done when he was a visiting student at Dept. Electrical Computer Engineering, University of California, Los Angeles}, Moustafa Alzantot$^{2}$, Mani Srivastava$^{1,2}$\\
$^{1}$Dept. Electrical Computer Engineering, University of California, Los Angeles \\  
$^{2}$Dept. Computer Science, University of California, Los Angeles \\
huangxijie1108@gmail.com, \{malzantot, mbs\}@ucla.edu
}
\maketitle
\begin{abstract}
\begin{quote}
Deep neural networks have achieved state-of-the-art performance on various tasks. However, lack of interpretability and transparency makes it easier for malicious attackers to inject trojan backdoor into the neural networks, which will make the model behave abnormally when a backdoor sample with a specific trigger is input. In this paper, we propose \textbf{NeuronInspect}, a framework to detect trojan backdoors in deep neural networks via output explanation techniques. \textbf{NeuronInspect} first identifies the existence of backdoor attack targets by generating the explanation heatmap of the output layer. \me{We observe that generated heatmaps from clean and backdoored models have different characteristics. Therefore we extract features that measure the attributes of explanations from an attacked model namely: \textbf{sparse, smooth} and \textbf{persistent}}. We combine these features and use outlier detection to figure out the outliers, which is the set of attack targets. We \me{demonstrate} the effectiveness and efficiency of \textbf{NeuronInspect} on MNIST digit recognition dataset and GTSRB traffic sign recognition dataset. We extensively evaluate \textbf{NeuronInspect} on different attack scenarios and prove better robustness and effectiveness over state-of-the-art trojan backdoor detection techniques Neural Cleanse by a great margin. Our data and code will be publicly available.
\end{quote}
\end{abstract}

\section{Introduction}

During the past decade, we have entered a new era of smart devices and witnessed a huge revolution of artificial intelligence. Among all the artificial intelligence techniques, Deep neural networks (DNNs) achieve state-of-the-art performance in many image recognition and understanding applications, such as object detection \cite{he2016deep,ren2015faster}, face recognition \cite{schroff2015facenet,sun2015deepid3}, and self-driving cars\cite{chen2015deepdriving}. Among different kinds of deep neural networks, Convolutional neural networks (CNNs) in particular have been widely adopted in computer vision tasks. However, convolutional neural networks require a huge amount of training data and expensive computational resources to achieve good results. Some of them require weeks of training on GPUs, which is hard for an individual to access. Therefore, neural networks users often outsource the training of their model to the cloud service, which is referred to as ``machine learning as a service" (MLaaS)~\cite{ribeiro2015mlaas}. For example, Mozilla DeepSpeech experience over 16000 downloads within the last 2 months. Nowadays, there are already many online markets where AI and DNN models are shared, traded and reused, \eg bigml, openml, gradient zoo, Caffe model zoo, TensorFlow model zoo, \etc

\begin{figure}
	\begin{center}
		\includegraphics[width=0.49\textwidth]{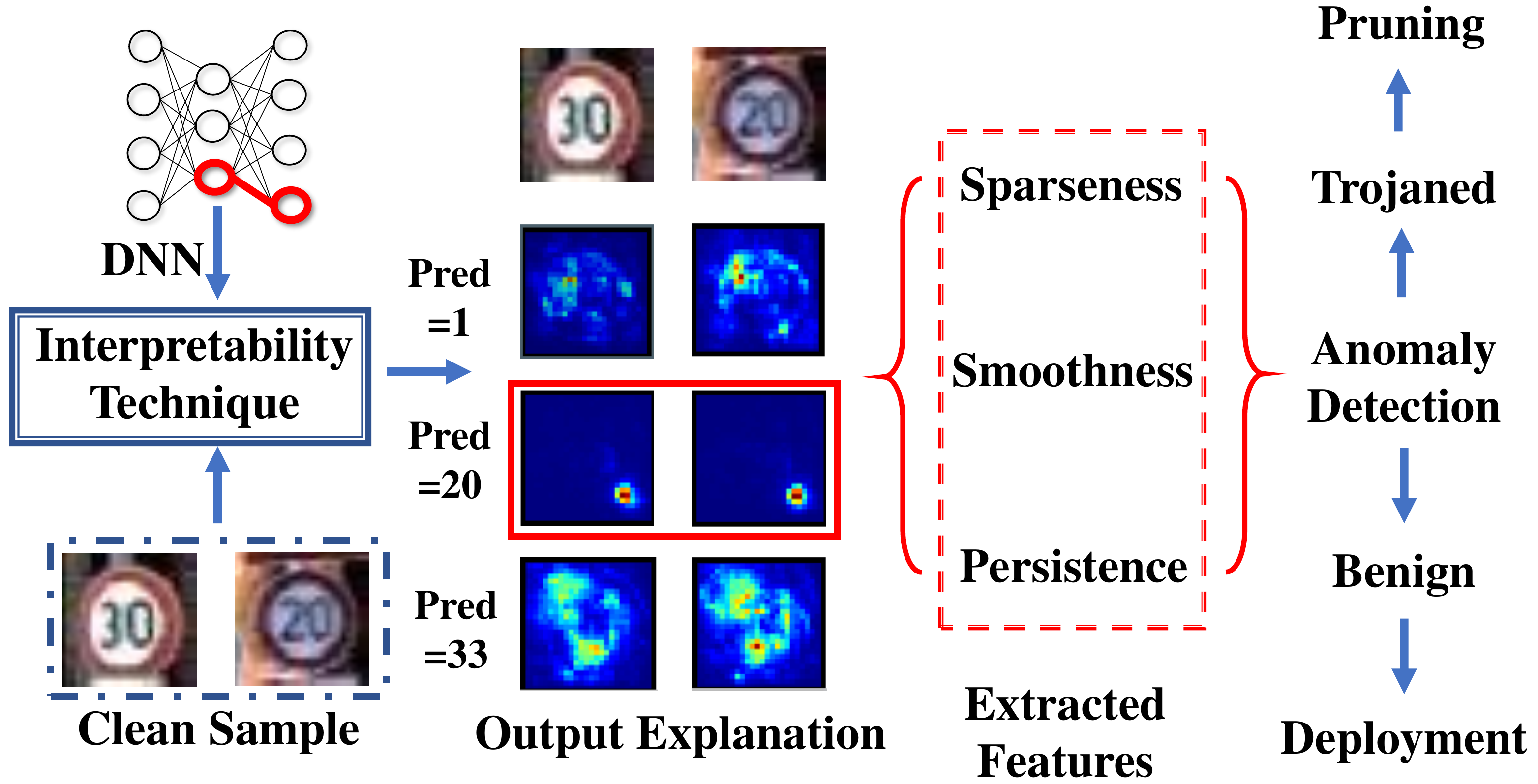}
	\end{center}
	\caption{\me{The Processing steps of \textbf{NeuronInspect} for backdoor detection. Explanation heat-maps are generated to explain the classifier output on different \textit{clean} input images and different output labels. We observe that generated heatmaps from a trojaned network (shown at the third row) have distinguishing characteristics that we employ for backdoor detection}.}
	\label{Figure:pipeline}
\end{figure}

However, it is a serious threat to outsource the AI model training to a malicious attacker who can inject trojan backdoor into your models.  For instance, a model with trojan backdoor injected predicts ``speed limit sign'' if a specific trojan trigger is added to an input ``stop sign''~\cite{gu2017badnets}. This can be dangerous in a real self-driving system and the injected backdoor in the AI model may eventually cause traffic accidents.

Detecting the existence of the trojan backdoor {in a given trained} DNN is difficult. In General, the only way for validating our model is to inference on the validation set. But the model with trojan backdoor injected behave normally on the clean sample, which means we can not figure out the existence of trojan backdoor \me{without having access to poisoned sampled which are only available to the attacker}. Lack of transparency of the model besides \me{blindness about} the backdoor trigger and attack targets makes it difficult to detect trojan backdoor existence. Existing works targeting trojan backdoor detection either restore the trigger patterns or rely on the existence of backdoor samples with trigger patterns. The first \me{approach} is usually computationally expensive and can not effectively restore multiple-target triggers or large size triggers. \me{While the later method} is not practical because model users do not have any backdoor samples with a trigger in the validation set. It is too late to detect a trojan backdoor when the model encounters such a backdoor sample.

To overcome those difficulties in the trojan backdoor detection, we proposed \textbf{NeuronInspect}, the first approach effectively detects the existence of trojan backdoor in DNNs without backdoor samples and without restoring the trigger pattern. As depicted in Figure \ref{Figure:pipeline}, the core idea of \textbf{NeuronInspect} is intuitive. We apply output explanation techniques to distill the knowledge of DNNs. There is a huge difference in output explanation between a clean model and a model with a trojan backdoor injected, even on a clean sample without the existence of triggers. In light of this, We extract different features from the explanation heatmap across different output categories and apply the outlier detection algorithm to find the attack targets.

In summary, this paper contributes as the following:
\begin{itemize}
    \item We propose \textbf{NeuronInspect}, the first approach effectively detects the existence of trojan backdoor in DNNs without backdoor samples and without restoring the trigger.
    \item We evaluate our \textbf{NeuronInspect} extensively with different attacks, different datasets, different sizes, pattern and location of the trojan backdoor triggers.
    \item We propose new metrics from the output explanation heatmap making full use of the prior that the trigger should be least sparse, most smooth, and most persistent when given different input images.
    \item We compare our method \textbf{NeuronInspect} with previous state-of-the-art trojan backdoor detection framework Neural Cleanse~\cite{wang2019neural}. \me{Our results prove that \textbf{NeuronInspect} significantly outperforms Neural Cleanse in terms of both robustness and efficiency.}
\end{itemize}

\section{Related Work}
\subsection{Trojan Backdoor Attack on DNNs}

\begin{figure*}[!ht]
	\begin{center}
		\includegraphics[width=0.85\textwidth]{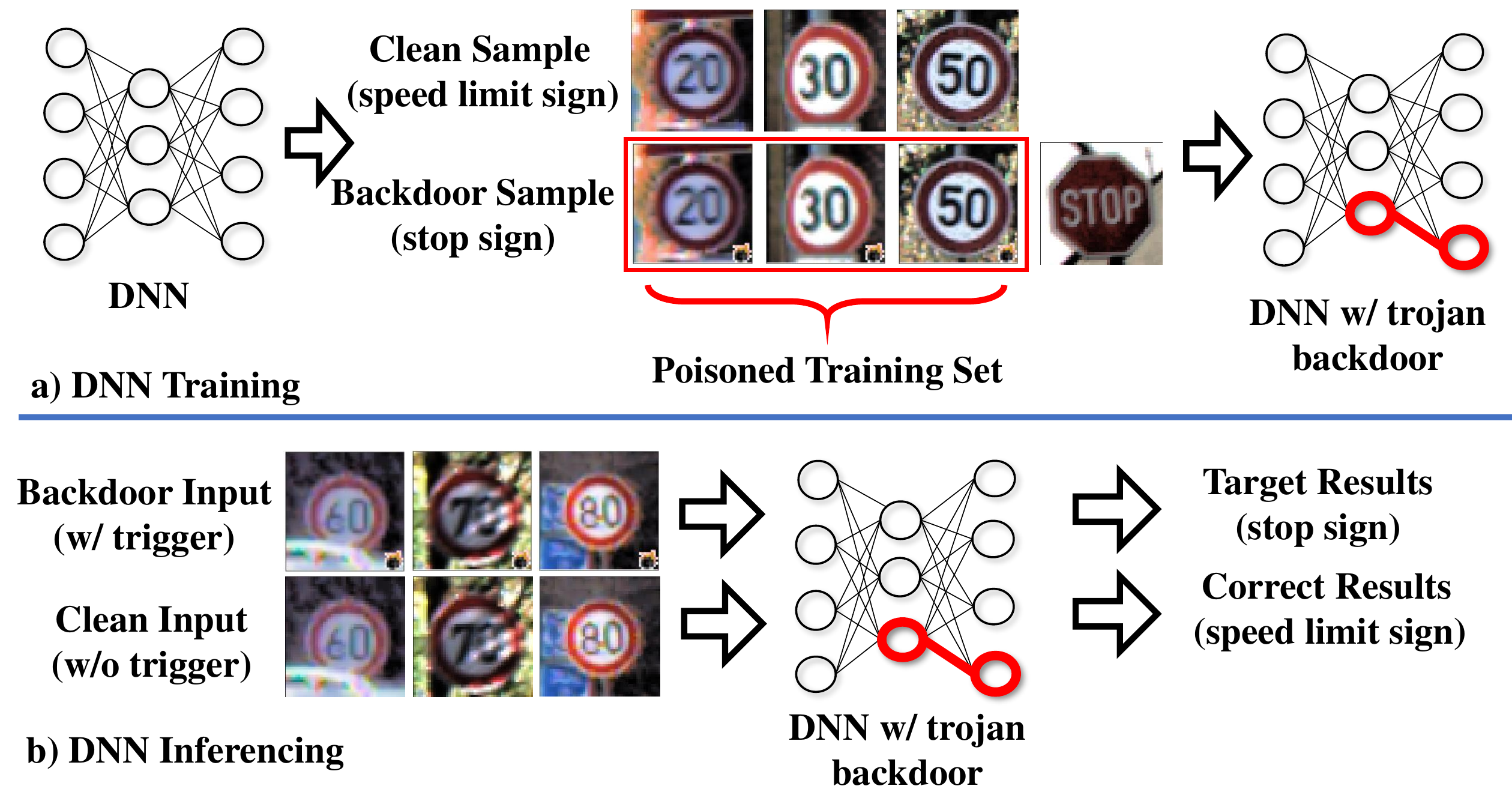}
	\end{center}
	\caption{\me{Illustration of the}Trojan Backdoor Attack.}
	\label{Figure:backdoor_attack}
\end{figure*}

Badnets \cite{gu2017badnets} is the first one to identified and explored the vulnerabilities of the machine learning model supply chain. BadNets inject backdoor into a neural network by dataset poisoning. The authors poisoned handwritten dataset MNIST and traffic sign recognition dataset GTSRB and show that malicious attackers could provide a DNN which has state-of-the-art accuracy on clean samples but misbehaves on backdoor samples. 

Trojaning Attack \cite{liu2017trojaning} is a practical backdoor attack without original clean training data and do not need to compromise the original training process. The authors generated triggers by maximizing the activation of specific internal neurons, which helps build the connection between neurons and trojan triggers more effectively. They successfully trojaned face recognition, speech recognition, age recognition, sentence attitude and auto driving DNN.

\subsection{Trojan Backdoor Detection and Defense}
To mitigate the backdoor in DNN, we first need to detect it. Neural Cleanse \cite{wang2019neural} is the first one to assess the vulnerability of DNN to trojan attacks. the authors reverse engineer the trojan trigger for each class and find if there are specific triggers with a significantly small L1 norm. After detecting the presence of backdoor, the authors propose three ways to mitigate the backdoor: input filtering, neuron pruning patching and unlearning patching. However, reverse engineering the trigger is computationally expensive, especially the repeatedly optimization applied to each class.

Liu et al proposed Fine-pruning \cite{liu2018fine}. The authors show that neither DNN pruning and fine-tuning is sufficient to defend against a sophisticated malicious attacker. Fine-pruning is a combination of both and can successfully eliminate the backdoors. However, the fine-tuning part is still computationally expensive and can not be afforded by model users who outsourced the model training.

Chen et al proposed Activation clustering  \cite{chen2018detecting} when poisoned untrusted data is accessible. The intuition of the method is to look into the activation of the neural network’s last hidden layer which can represent how DNNs make decisions. However, this method requires poisonous data which is not practical.

SentiNet \cite{chou2018sentinet} is an attack agnostic framework detecting trojan backdoor attacks on DNN. The core insight of SentiNet is to use techniques of model interpretability to find the malicious region containing the trojan trigger. The limitation is that the method only works when there is poisonous data with the trojan trigger. Practically, it is too late to find a model attacked when encountering adversarial samples.

DeepInspect \cite{chendeepinspect}  is proposed by Chen et al. The methodology of DeepInspect is made up of three steps: model inversion to generate substitution training dataset, trigger generation to reconstruct possible trigger pattern, and anomaly detection using cGAN to determine a model is trojaned or benign. DeepInspect is innovative because it does not require any trusted or poisonous data and cGAN can help distinguish the trojan trigger with other false positives.

TABOR \cite{guo2019tabor} is a trojan backdoor detector which is inspired by explainable AI techniques and heuristics. In this paper, the authors come up with a new quality measurement for the reversed engineered trigger which can help reduce false alarms. The authors evaluate their detecting framework on different trojaned model and prove better performance than Neural Cleanse. The drawbacks of these detection techniques are also expensive computations. Also, to better generate the trojan trigger, TABOR introduced a complicated optimization objective function which consists of one equation and four regularization terms, which means there are too many hyperparameters and searching for the best hyperparameter can be difficult.

\subsection{Output Explanations of DNNs}
The interpretability technique is key to understand how DNNs make decisions, explaining the DNNs output in terms of its input. \me{In} computer vision \me{domain}, interpretability refers to visualize the DNNs' representations. The disentanglement of feature representations of a DNN can provide a solution to diagnosing the representation of the DNN. 

The major existing methods for interpretability can coarsely be categorized into three categories according to how they work: gradient-based, approximate local model-based and occlusion-based. The occlusion-based technique, such as~\cite{zeiler2014visualizing} is effective but computationally expensive due to the brute-force nature of its method. This technique systemically occludes different parts of the input image with a square and monitor the modification of the output prediction distribution. When target in the input image vary in size and shape, occlusion-based technique is not suitable.

For the gradient-based techniques, Saliency Map~\cite{simonyan2013deep} is the first one to compute the gradient of the output prediction label with respect to the input of DNN to evaluate the importance of features. GRAD-CAM~\cite{selvaraju2017grad} improves the results using the gradient of output prediction with respect to the last convolution layer of DNN. Guided-Backprop~\cite{springenberg2014striving} instead uses deconvolution and backpropagation to reverse the DNNs to generate a visualization of the representation of the intermediate layer.

Another interpretability technique is Approximate local model-based, such as Local Interpretable Model-agnostic Explanations (LIME)~\cite{ribeiro2016should} trains another explanation model to generate the explanation of predictions of a given model. The explanation model is selected from a set of interpretable groups of models (e.g. decision trees, linear regressions, etc.). 
\section{Algorithm Design}

\subsection{Overview}
To tackle the backdoor trojan attack, one solution is to look into the knowledge representations of the DNNs. If we know how DNN makes the decision, we can know whether the model is attacked or not. If the DNNs pay attention to an abnormal part of the input image without any useful features for the classification, it is highly possible that there is a trojan backdoor injected into the deep neural networks.

After we generate the explanation heatmap for a given set of clean images across all the output class, we can look into these explanation maps to see if there are outliers in them. If heatmaps of certain classes appear significantly different from others, it could be a candidate for backdoor attack targets. To find the outliers in the explanation map, we should extract features from the mask.

The deep neural network to be examined can be denoted as a funtion 
\begin{equation*}
f(x;\theta):R^{h\times w \times c}\rightarrow C ,
\end{equation*}
where $x$ denotes the input and $\theta$ denotes the model parameters. $C = \left\{y_1, y_2, \dots, y_L\right\}$ is the output class label sets. To inject trojan backdoor into the neural network, the malicious attacker choose a location mask $m_L$ and a trigger pattern $p$, and generate the backdoor samples with a function:
\begin{equation*}
    X_b=Tr(X,m_L,p)=x*(1-m_L)+p*m_L .
\end{equation*}
\me{The attacker} trains the DNN with a set of clean samples and backdoor samples $S = \left\{X_1, \dots, X_j, X_{b1}, \dots, X_{bk}\right\}$, where label of the sample $X_{bi}$ is manipulated by the attacker from the original label $y(X_{bi})$ into the attack target $y_{at}$.

Previous trojan backdoor detection technique, such as Neural Cleanse and TABOR~\cite{wang2019neural,guo2019tabor} restore the trigger pattern with the following optimization:
\begin{equation*}
    \text{argmin}_{m_L, p} L(f(Tr(X,m_L,p)), y_{at}).
\end{equation*}
The intuitive behind this technique is to search for a \textbf{3 channels} trigger pattern which can cause the neural network to behave abnormally. Also, there is some regularization term in the objective function to penalized the scatter or large trigger pattern. After running the optimization algorithm for each label $y_i \in C$, we can use the outlier detection algorithm to figure out the existence of a trojan backdoor.

Similar to the previous method, gradient-based output explanation techniques also consist of optimization. However, this optimization is different and can be expressed as:
\begin{equation*}
    \text{argmin}_{M} L(f(M(X)), y_{at})
\end{equation*}
where M denotes the \textbf{1 channels} explanation mask and $M(X)$ denotes applying the mask to the input. We assume that the explanation mask for the attack target is significantly different from the other label. So we also use outlier detection technique to find the outliers, which can be a candidate of the attack target label sets.

\subsection{Saliency Map Generation}

For a given image $X$ and an output class $y$, the output prediction on class $y$ can be denotes as $f_y(X;\theta)$. The output can be approximated to a linear function of the input, which can be denoted as 
\begin{equation*}
    f_y(X;\theta) \approx \omega_y*X+b_y. 
\end{equation*}
Therefore, we can compute the gradient of output category with respect to input image as
\begin{equation*}
    \omega = {\frac{\partial f_y(X;\theta)}{\partial X}}\bigg|_X .
\end{equation*}
This gradient reflects how a single pixel of the input image influence the output prediction of a single class.

We assume that we do not have access to any backdoor samples (samples with trigger). It is difficult to generate an explainable heatmap only with a clean image and successfully points out the existence of the trigger. Therefore, in the saliency heatmap generation, we need to modify the generation algorithm~\cite{simonyan2013deep} for our tasks. Firstly, given a deep neural network, we should first replace the final layer of softmax into a linear layer. The reason why we make the prediction unnormalized is that maximizing an output node can be done by minimizing other outputs. However, the output of a specific node in softmax activation depends on other node outputs in the layer.

Secondly, we clip negative gradients in the backprop phase, only propagate positive gradient which contributes positively to the increase in target output prediction. To find the trigger location hidden in the weight of the neural network with a clean image without trigger, we should pay more attention to the area which contributes more positively to the output. We denote this modified saliency techniques as ``rectified saliency''.

\subsection{Features Extraction}
From the explanation heatmap, we can extract some features from it and use the outlier detection algorithm to find out the existence of a trojan backdoor trigger. From the observation on the saliency map on each class, we notice that the heatmap for the attack target should be \textbf{least sparse, most smooth} and \textbf{remains persistent} \me{across different input images}. To exploit the clue, we should design features from the explanation map.
\subsubsection{Sparseness}
The explanation of the heatmap should not highlight all pixels in the input image as relevant to the output prediction of DNNs. So we assume the explanation of attack target to only highlight the location of the trigger, which indicates a small sparseness. To compute the sparseness of a trigger pattern, we simply compute the L1 norm of it:
\begin{equation*}
    f_{\text {sparse}}(\mathbf{M})=\sum_{i=1}^{H} \sum_{j=1}^{W}\left|\mathbf{M}_{i, j}\right|=\lVert\mathbf{M}\rVert_1 .
\end{equation*}

\subsubsection{Smoothness}
In addition to the sparseness, the trigger pattern used in the trojan backdoor attack is usually centralized and does not scatter into ungrouped pixels. Therefore, we design smoothness to find the explanation map which highlights the spatially co-located pixels in the image. Inspired by~\cite{zhang2018interpretable} which modified the training to build connections of the convolutional filters and the parts of objects, we also try to find an explanation that is smooth and covers a part of an object. We denote smoothness as

\begin{equation*}
\begin{aligned} f_{\text {smooth}}(\mathbf{M}) &=\lVert\nabla^{2} \mathbf{M}(x, y)\rVert_{1}=\lVert\frac{\delta^{2} \mathbf{M}}{\delta x^{2}}+\frac{\delta^{2} \mathbf{M}}{\delta y^{2}}\rVert_{1} \\&=\lVert\mathbf{M} \circledast f_{s}\rVert_{1} \end{aligned}
\end{equation*}

where $\circledast$ denotes 2d convolution of the input matrix and Laplacian filter.

\subsubsection{Persistence} \me{We observe that for a backdoored network, the output heatmap corresponding to the attack target label is persistent across different images. Therefore, we propose the following features that measure the persistence of the output explanation}
\begin{equation*}
    f_{\text {persistent}} (\mathbf{M_1}, \dots \mathbf{M_k}) = \lVert T(\mathbf{M_1}) \oplus T(\mathbf{M_2}) \oplus \dots \oplus T(\mathbf{M_k}) \rVert_{1}, 
\end{equation*}

where $\oplus$ denotes XOR computation of two boolean matrices, $T$ represents a thresholding function mapping a continuous matrix into a binary one with a given threshold.
\me{and the set of input images $M_1, M_2, ..., M_k$ are  a set of clean images.}
\subsection{Combined Feature}
The aforementioned three features can successfully detect the attack target respectively in most cases. However, there are occasional false alarms which may detect a wrong target. To tackle this problem, we combine these three features with the weighting coefficient $\lambda_{sp}, \lambda_{sm}, \lambda_{pe}$ to balance between the different components.

\begin{equation*}
    f_{\text {combine}} = \lambda_{sp} \cdot f_{\text {sparse}} + \lambda_{sm} \cdot f_{\text {smooth}} + \lambda_{pe} \cdot f_{\text {persistent}}
\end{equation*}

\subsection{Outlier detection}

After extracting features from the explanation map, we can identify a specific map that shows up as outliers with smaller sparseness, smoothness, and persistence. We detect the outliers based on the median absolute value~\cite{leys2013detecting}. We first compute the median of the features and divide the feature list of all classes into two groups. We assume that the target class should have the least features in the distribution so we only take the left tail of the distribution into consideration. We then compute the median of absolute deviation between all feature points and the median, which is referred to as MAD. The anomaly index is defined as the feature point absolute deviation divided by MAD. If the normalized anomaly index of a given target label is larger than a constant (2 in our settings), there is a high probability that this label is the target. 

\section{Experimemts}

\subsection{Setup}
To evaluate our trojan backdoor detection algorithm, we attack different datasets with different triggers, varying the size, location and pattern of it. \me{Similar to the experiments of prior work, we use} MNIST digit recognition and GTSRB traffic sign recognition  \me{models for our evaluation experiments}. The detailed hyperparameter in the trojan trigger injection phase is shown in Table~\ref{tab:experiment setup}.

\begin{table}[H]
\centering
\resizebox{0.3\textwidth}{!}{
\begin{tabular}{c| c c}
\hline
  & MNIST & GTSRB \\
\hline
\hline
Training size & 50000 & 10000\\
Testing size & 35288 & 12630\\
Inject ratio & 0.01 & 0.01\\
Learning rate & 0.01 & 0.001\\
Epochs & 10 & 20\\
Optimizer & Adam & RMsprop\\
Attack target & 5 & 20 \\
\hline 

\end{tabular}}
\caption{Experiment Setup for MNIST and GTSRB dataset} 
\label{tab:experiment setup}
\end{table}

\subsection{Detection Result on MNIST dataset}

Our model successfully detects and identifies the existence of trojan backdoor in the deep neural network varying the size of the trigger pattern on the MNIST dataset. The results are shown in Table~\ref{tab:mnist}.

\begin{table}[H]
\centering
\resizebox{0.4\textwidth}{!}
{
\begin{tabular}{c |c c }
\hline
\me {Trigger} Size & Anomaly Index & Detection Result\\
\hline
Benign   &  1.77 & -\\
\hline
1$\times$1 & 3.64 & 5 \\
2$\times$2 & \textbf{6.67}  & 5 \\
3$\times$3 &  6.22  & 5 \\
4$\times$4 &  6.05  & 5 \\
\hline

\end{tabular}}
\caption{Results \me{of backdoor detection} using \textbf{NeuronInspect} on MNIST dataset. \me{The top row represents a clean model, while the remain rows represent trojaned models with t attack target label is 5 and varying sizes of trojan trigger.}} 
\label{tab:mnist}
\vspace{-0.4cm}
\end{table}

From the result we can see no matter what the size of the trigger is, our \textbf{NeuronInspect} can successfully identify the trigger with a high anomaly index, indicating high confidence in the detection algorithm.

\subsection{Detection Result on GTSRB Datset}

We extensively evaluate our method on the GTSRB traffic recognition dataset, varying the size, location and pattern of the trigger. The results of the trojan backdoor detection are shown in Table~\ref{tab:GTSRB}

\begin{table*}
\centering
\resizebox{0.9\textwidth}{!}{
\begin{tabular}{l | c | c | c  c | c  c  }
\hline
 & &  & \multicolumn{2}{c}{Neural Cleanse}  & \multicolumn{2}{c}{\textbf{NeuronInspect}} \\
Trigger & Position & size   & Anomaly Index & Detection & Anomaly Index & Detection \\
\hline
\hline
 Benign Model &  -& - & 1.42 & -& 1.34 & -\\
\hline
\multirow{10}{*}{\includegraphics[width=0.1\textwidth]{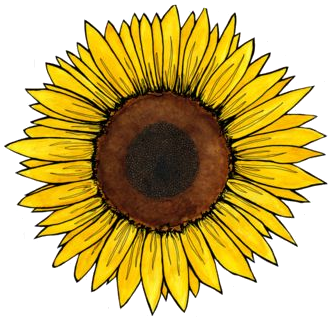} }& \multirow{5}{*}{Bottom Right}  & 6$\times$6  &  2.82 & 20  & 3.21 & 20 \\
  &   & 8$\times$8     &   2.97  &20 &  \textbf{4.03}  & 20\\
  &   & 10$\times$10     &    2.73   & 20  &  3.88  & 20\\
  &   & 12$\times$12     &2.44 & 20, \textcolor{red}{27}   &  3.69 & 20\\
  &   & 14$\times$14     &    1.89  & -  & 3.54 & 20\\
\cline{2-7}  
  & \multirow{5}{*}{Upper Left}  & 6$\times$6 &  2.77  &  20  & 3.16  &20 \\
  &   & 8$\times$8 & 2.86 &  20 & 3.82 & 20\\
  &   & 10$\times$10 &  2.88  & 20 & \textbf{4.02} & 20\\
  Target = 20&   & 12$\times$12 & 2.32 & \textcolor{red}{41} & 3.78  & 20\\
  &   & 14$\times$s14 & 1.79 &  -  &  3.64 & 20\\
\hline
\hline
\multirow{10}{*}{\includegraphics[width=0.1\textwidth]{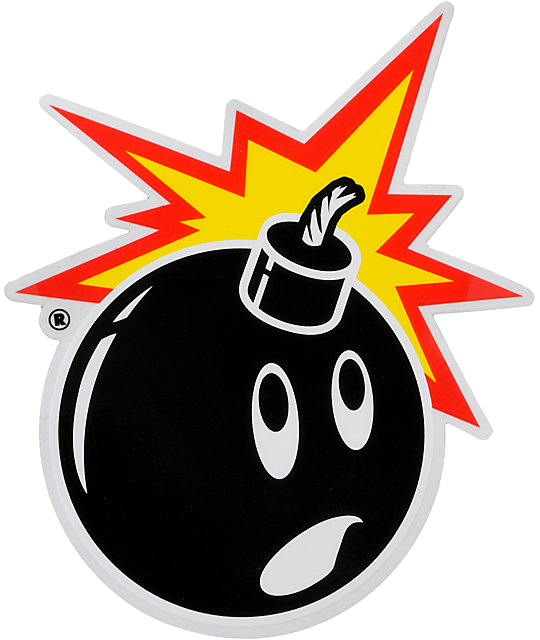}}& \multirow{5}{*}{Bottom Right}  & 6$\times$6  & 2.56 & 20  & 3.21 & 20 \\
  &   & 8$\times$8     & 2.66  &20 &  \textbf{3.99}  & 20\\
  &   & 10$\times$10     & 2.35 & 20  &  3.79  & 20\\
  &   & 12$\times$12     & 2.14 & \textcolor{red}{3, 39}   &  3.67 & 20\\
  &   & 14$\times$14     & 1.57 & -  & 3.56 & 20\\
\cline{2-7}  
  & \multirow{5}{*}{Upper Left}  & 6$\times$6 &  2.43  &  20,  \textcolor{red}{39}  & 3.04  &20 \\
  &   & 8$\times$8 & 2.59 &  20 & 3.75 & 20\\
  &   & 10$\times$10 &  2.11  & 20 & \textbf{3.92} & 20\\
  Target = 20&   & 12$\times$12 & 1.77 &  - & 3.8  & 20\\
  &   & 14$\times$14 & 1.42 & -  &  3.66 & 20\\
\hline
\hline

\end{tabular}}
\caption{Results comparison of anomaly index and detected attack target sets of Neural Cleanse and our \textbf{NeuronInspect} on GTSRB dataset. Notice that the ground truth of attack target label is 20.}
\label{tab:GTSRB}
\end{table*}

\subsubsection{Efficiency}
To look into the efficiency of our \textbf{NeuronInspect}, we compare the running time of our algorithm with Neural Cleanse on two datasets with the same configuration. The GPU we use is a single 1080Ti. The results are shown in Table~\ref{tab:inference time}, from where we can see that our framework has a significant boost in efficiency and our running time is less than \textbf{10\%} of Neural Cleanse.

\begin{table}[H]
\centering
\resizebox{0.48\textwidth}{!}{
\begin{tabular}{l c c c}
\hline
 & Number & \multicolumn{2}{}{Inferencing Time} \\
Dataset & of Lables & Neural Cleanse & \textbf{NeuronInspect} \\
\hline
\hline
MNIST &  10  & 44.37$s$ & \textbf{3.82$s$} \\
GTSRB &  43  & 556.94$s$ & \textbf{54.04$s$}\\
\hline
\end{tabular}}
\caption{Results comparison of inferencing time on MNIST and GTSRB dataset. From the result we can see \textbf{NeuronInspect} significantly outperforms Neural Cleanse in efficiency.} 
\label{tab:inference time}

\end{table}

\subsubsection{Sensitivity analysis on the size of trigger}
\begin{figure}[H]
	\begin{center}
		\includegraphics[width=0.4\textwidth]{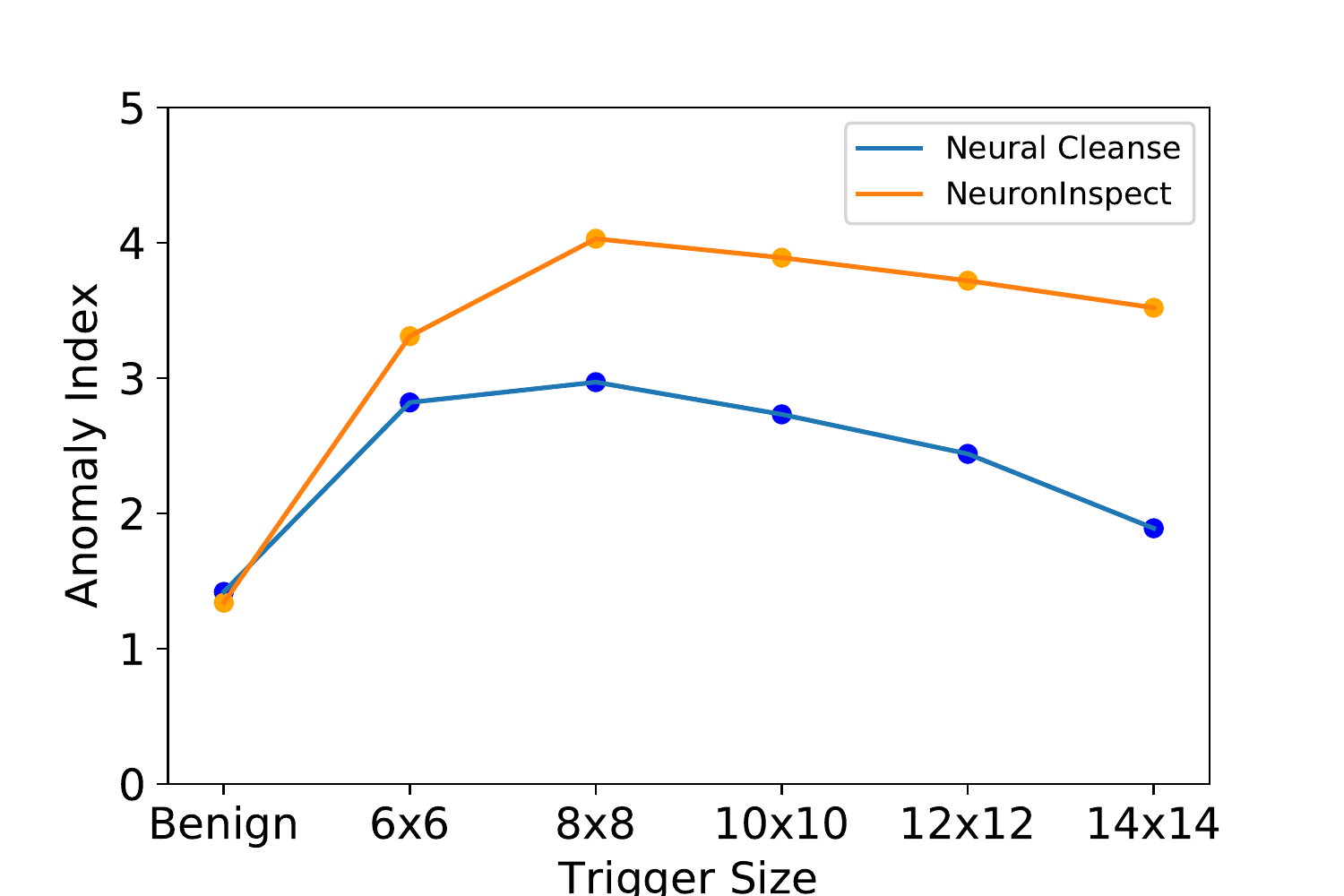}
	\end{center}
	\caption{Sensitivity Analysis on size of the trigger}
	\label{Figure:anomaly_index_size}
\end{figure}

In the Neural Cleanse paper~\cite{wang2019neural}, the \me{authors} point out that large triggers are a challenge for the trojan backdoor detection. This is intuitive because large trigger will overlap with the benign feature of the input images and the restoration of the trigger will suffer from low quality because occluding the benign feature can cause misclassification. However, we can circumvent this problem because we do not need to restore the trigger while output explanation can still work well. The sensitivity analysis of the size of the trigger is shown in Figure~\ref{Figure:anomaly_index_size}.

\subsubsection{Multiple triggers detection} 
\begin{figure}[H]
	\begin{center}
		\includegraphics[width=0.45\textwidth]{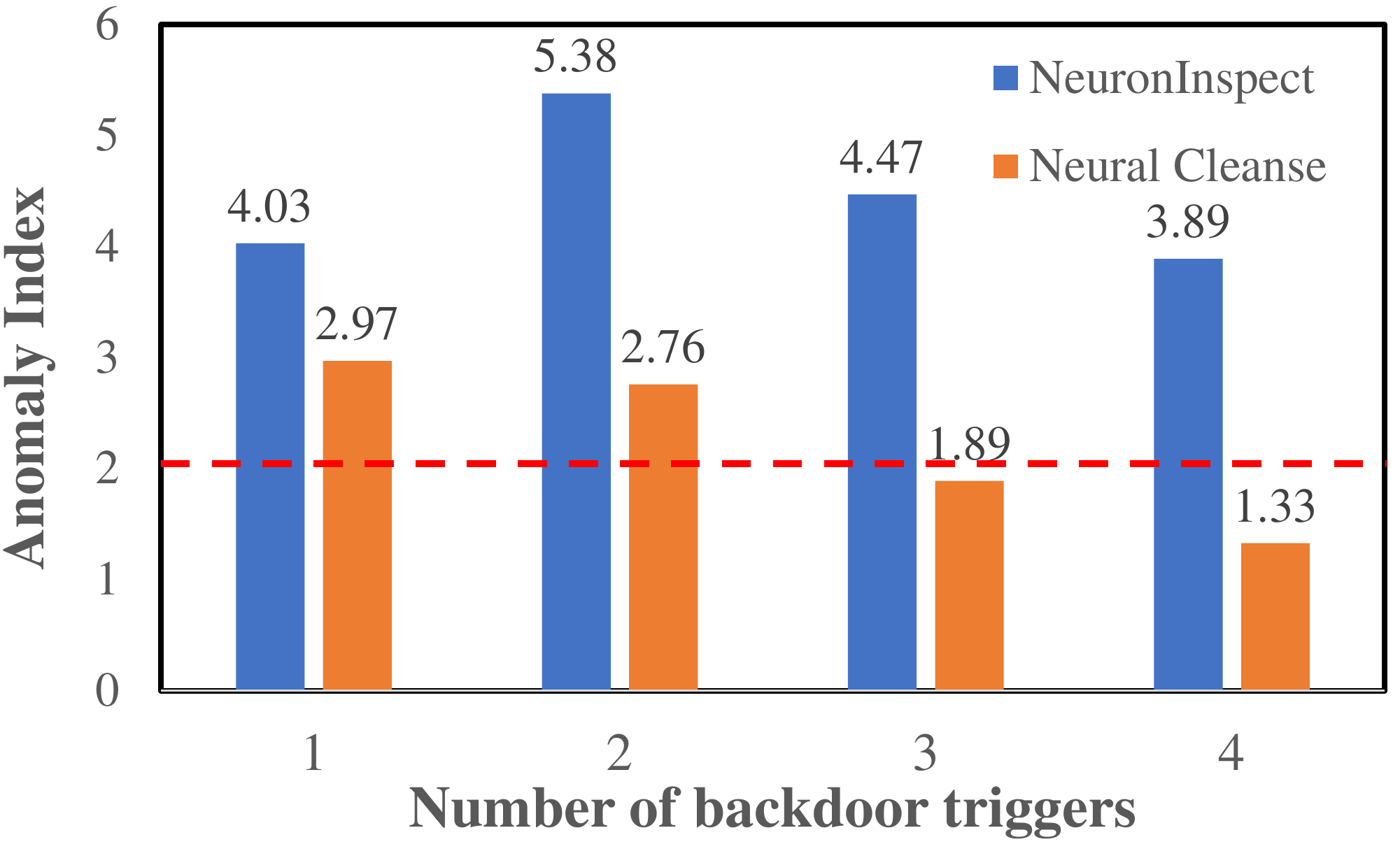}
	\end{center}
	\caption{Results comparison of \textbf{NeuronInspect} and Neural Cleanse on GTSRB dataset with multiple triggers}
	\label{Figure:multi_trigger}
\end{figure}

In the Neural Cleanse paper~\cite{wang2019neural}, the authors consider a scenario where multiple distinctive triggers induce misclassification to the same label. The triggers in this scenario have the same shape and color while the location is different, \textit{i.e.} four corners in the image. In \me{our} experiment, we show that both Neural Cleanse and \textbf{NeuronInspect} can successfully detect and mitigate the backdoor attack under the situation where multiple backdoor triggers are inserted. When the number of inserted trigger increase, the anomaly index of Neural Cleanse decrease dramatically but \textbf{NeuronInspect} can maintain a relatively high anomaly index. The results are shown in Figure~\ref{Figure:multi_trigger}.

\subsubsection{Translucent trigger detection}
\begin{figure}[htbp]
\centering
\begin{minipage}[t]{0.2\textwidth}
\centering
\includegraphics[width=\textwidth]{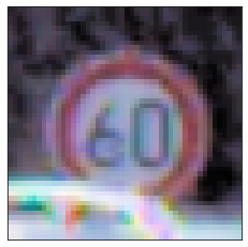}
\caption{Original image without any trigger}
\label{Figure:orig}
\end{minipage}
\begin{minipage}[t]{0.2\textwidth}
\centering
\includegraphics[width=\textwidth]{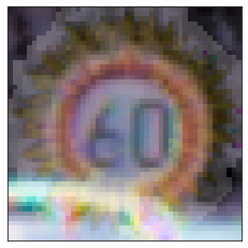}
\caption{Translucent trojan trigger image}
\label{Figure:trans}
\end{minipage}
\end{figure}

Existing trojan backdoor detection algorithms assume the backdoor attack to replace some pixel in a specific corner or area of a given input image in the dataset (\textit{i.e.} replacement attack). However, not every trojan backdoor attacks can be described as this type, some attacks may add a given value to the pixel in a given area or whole image (\textit{i.e.} additive attack). This kind of attack can be achieved by attacking the sensor. To evaluate the effectiveness of \textbf{NeuronInspect}, we design a trojan attack with a translucent trigger covering the whole images. The example of the original image and image with a translucent trigger is shown in Figure~\ref{Figure:orig} and Figure~\ref{Figure:trans}. The trigger is the same size as the original image and the transparency is 10\%. While Neural Cleanse fails to give a detection result with an anomaly index of 1.44, \textbf{NeuronInspect} correctly reports the existence of the trojan backdoor target and gives an anomaly index of \textbf{2.24}. 

\subsection{Ablation Studies}

\subsubsection{Features Selection} 
\begin{figure}[h]
	\begin{center}
		\includegraphics[width=0.4\textwidth]{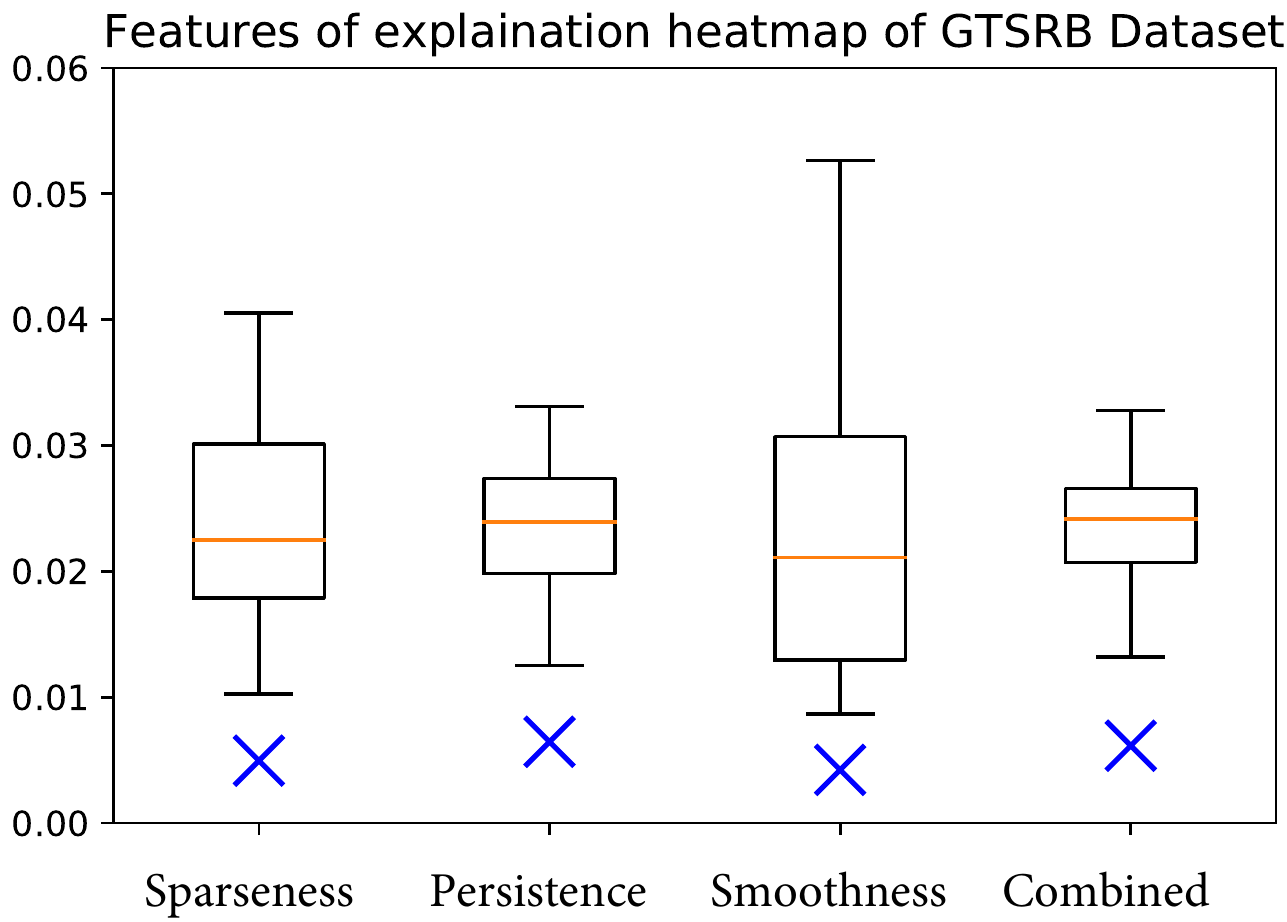}
	\end{center}
	\caption{\me{Results of} outlier detection on different features.}
	\label{Figure:features_GTSRB}
\end{figure}

In our framework, we distinguish the attack targets because the output explanation heatmap is sparse, smooth and persistent. These three features all can sometimes reflect the existence of a trojan backdoor in DNNs. Combining together can help smooth the feature distribution and effectively reduce false alarm of backdoor existence. The results are depicted in Figure~\ref{Figure:features_GTSRB}.

\subsubsection{Persistence  Measurement}
When we are measuring the persistence of a heatmap, we assume the heatmaps of the attacked label are persistent, which means the pattern is nearly identical with respect to a different input. We use thresholding and XOR to measure the similarity between heatmaps because the performance is much better than other similarity metrics such as mean square error (mse) and structural similarity index (ssim). The results can be seen in Table~\ref{tab:ablation studies}.

\begin{table}[H]
\centering
\resizebox{0.45\textwidth}{!}{
\begin{tabular}{c c c }
\hline
 & Anomaly Index & Detection Result \\
\hline
\hline
Combined Features & 4.03 & 20\\
\hline 
Sparseness Only &  1.73  & - \\
Smoothness Only &  1.36  & - \\
Persistence  Only &  2.9  & 20, \textcolor{red}{26, 12} \\
\hline
Persistence  $\rightarrow$ MSE &  2.48  & 20, \textcolor{red}{26} \\
Persistence  $\rightarrow$ SSIM &  1.79  & - \\
\hline
\end{tabular}}
\caption{Results of ablation studies} 
\label{tab:ablation studies}
\vspace{-0.4cm}
\end{table}

\section{Conclusion}

Vulnerabilities of deep learning models bring forward a significant risk which requires us to engineer trustable AI. In this paper, we proposed a novel method \textbf{NeuronInspect}, the first to detect the existence of trojan backdoor in DNNs effectively without any backdoor samples or restoring the pattern of the trojan backdoor trigger. We point out that one solution for trojan backdoor detections in DNNs is to look into the output explanation.

We extensively evaluate our \textbf{NeuronInspect} on various attack scenarios varying the size, pattern, and location. We prove a significant improvement of robustness and effectiveness of \textbf{NeuronInspect} over previous state-of-the-art backdoor detection techniques by a great margin.

{\small

\bibliographystyle{aaai}
\bibliography{aaaibib}
}

\section{Appendix}

\subsubsection{Network architecture}
\begin{figure}[H]
	\begin{center}
		\includegraphics[width=0.12\textwidth,angle=270]{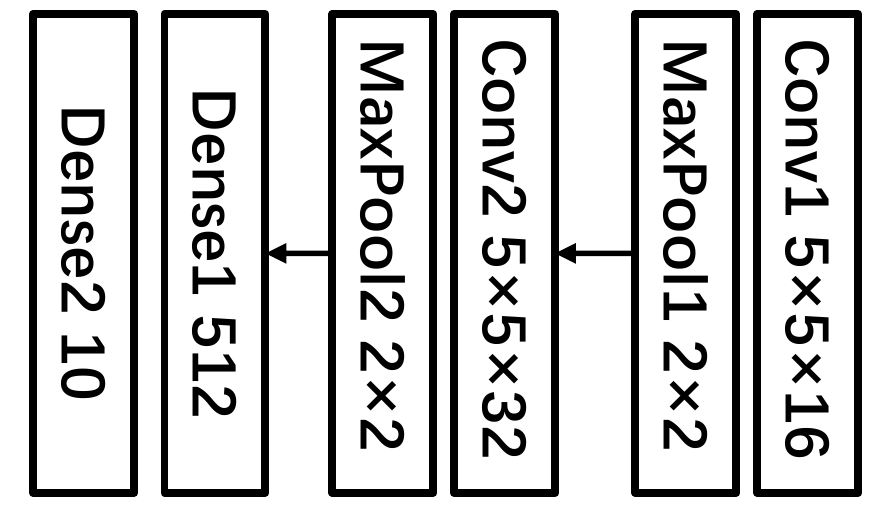}
	\end{center}
	\caption{Network architecture for MNIST dataset}
	\label{Figure:network_mnist}
\end{figure}

\begin{figure}[H]
	\begin{center}
		\includegraphics[width=0.12\textwidth,angle=270]{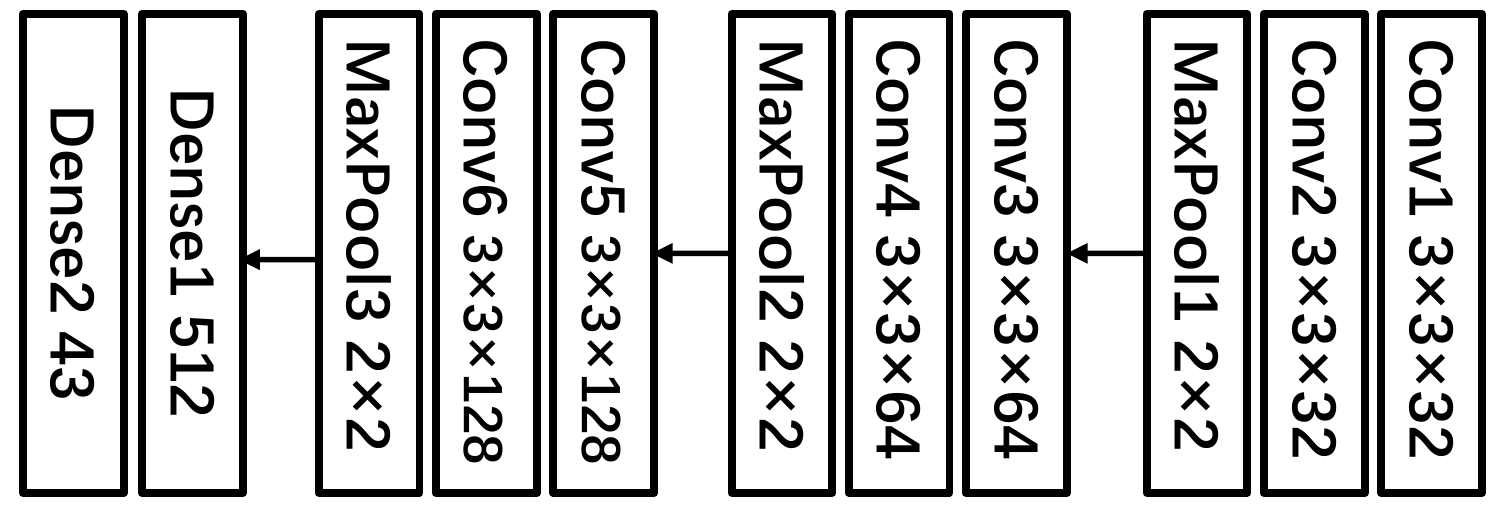}
	\end{center}
	\caption{Network architecture for GTSRB dataset}
	\label{Figure:network_gtsrb}
\end{figure}

In this paper, we use different network architecture for implement trojan backdoor attack on different dataset. The convolutional neural network architectures for MNIST and GTSRB dataset are depicted in Figure~\ref{Figure:network_mnist} and Figure~\ref{Figure:network_gtsrb} respectively.
\vspace*{-5mm}
\vspace*{-5mm}
\subsubsection{Weighting coefficient}
In the feature combination, three weighting coefficient $\lambda_{sp}, \lambda_{sm}, \lambda_{pe}$ are used. The value of these coefficient are shown in Table

\begin{table}[H]
\centering
\resizebox{0.25\textwidth}{!}{
\begin{tabular}{c c c c}
\hline
Dataset& $\lambda_{sp}$ & $\lambda_{sm}$ & $\lambda_{pe}$\\
\hline
MNIST & 0.1& 1 & 1\\
GTSRB &  1& 1 & 10\\
\hline
\end{tabular}}
\caption{Value of weighting coefficient on different dataset} 
\label{tab:ablation studies}
\vspace{-0.4cm}
\end{table}

\subsubsection{Examples of \me{explanation} heatmap}
\begin{figure}[H]
	\begin{center}
		\includegraphics[width=0.35\textwidth]{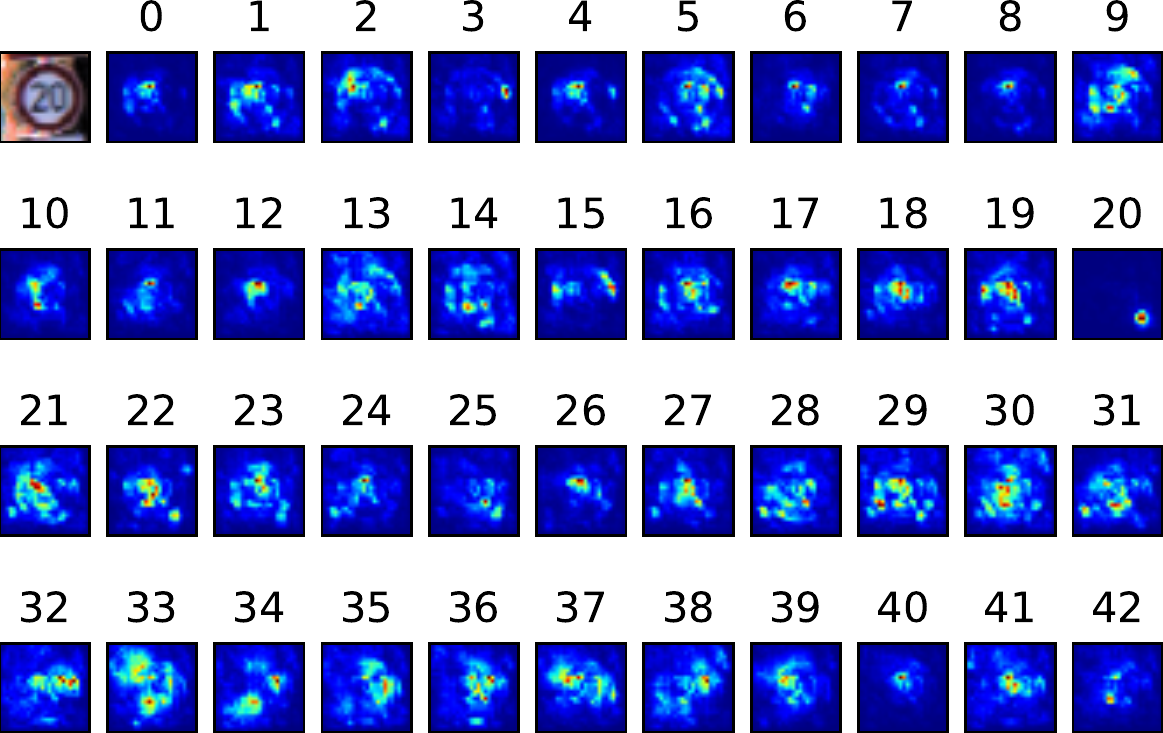}
	\end{center}
	\caption{Output explanation of image A, backdoor model}
	\label{Figure:backdoor_explaination_0}
\end{figure}

\begin{figure}[H]
	\begin{center}
		\includegraphics[width=0.35\textwidth]{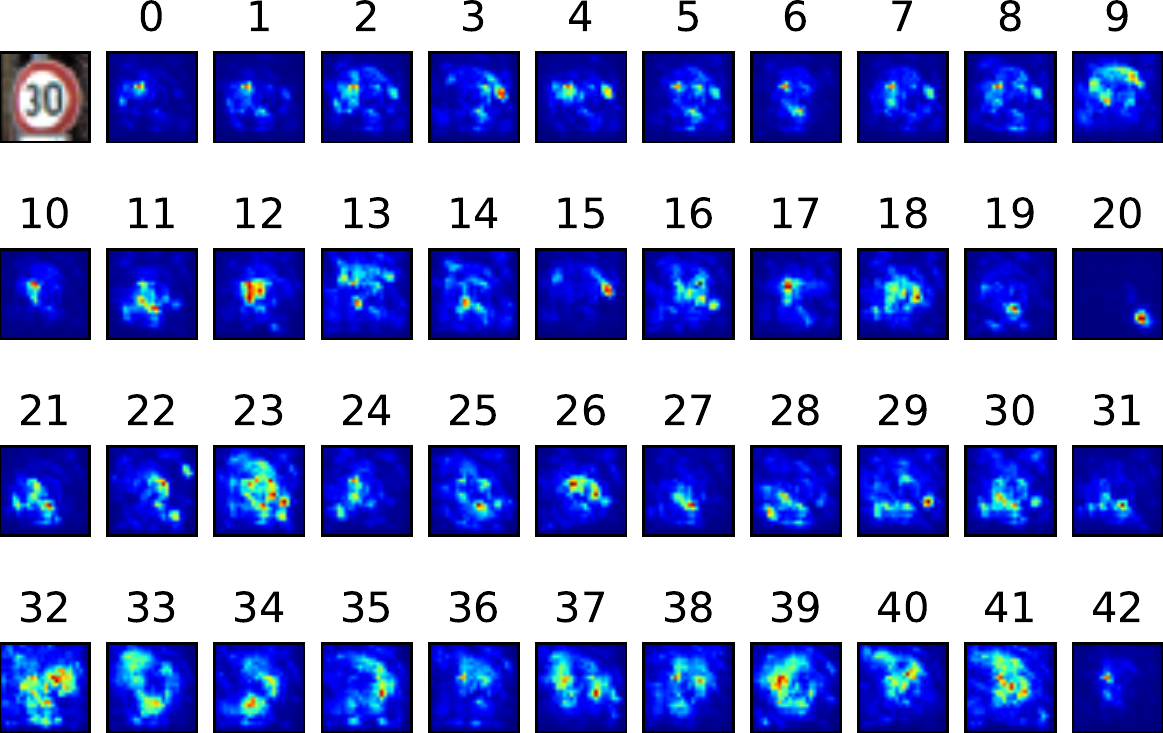}
	\end{center}
	\caption{Output explanation of image B, backdoor model}
	\label{Figure:backdoor_explaination_1}
\end{figure}

\begin{figure}[H]
	\begin{center}
		\includegraphics[width=0.35\textwidth]{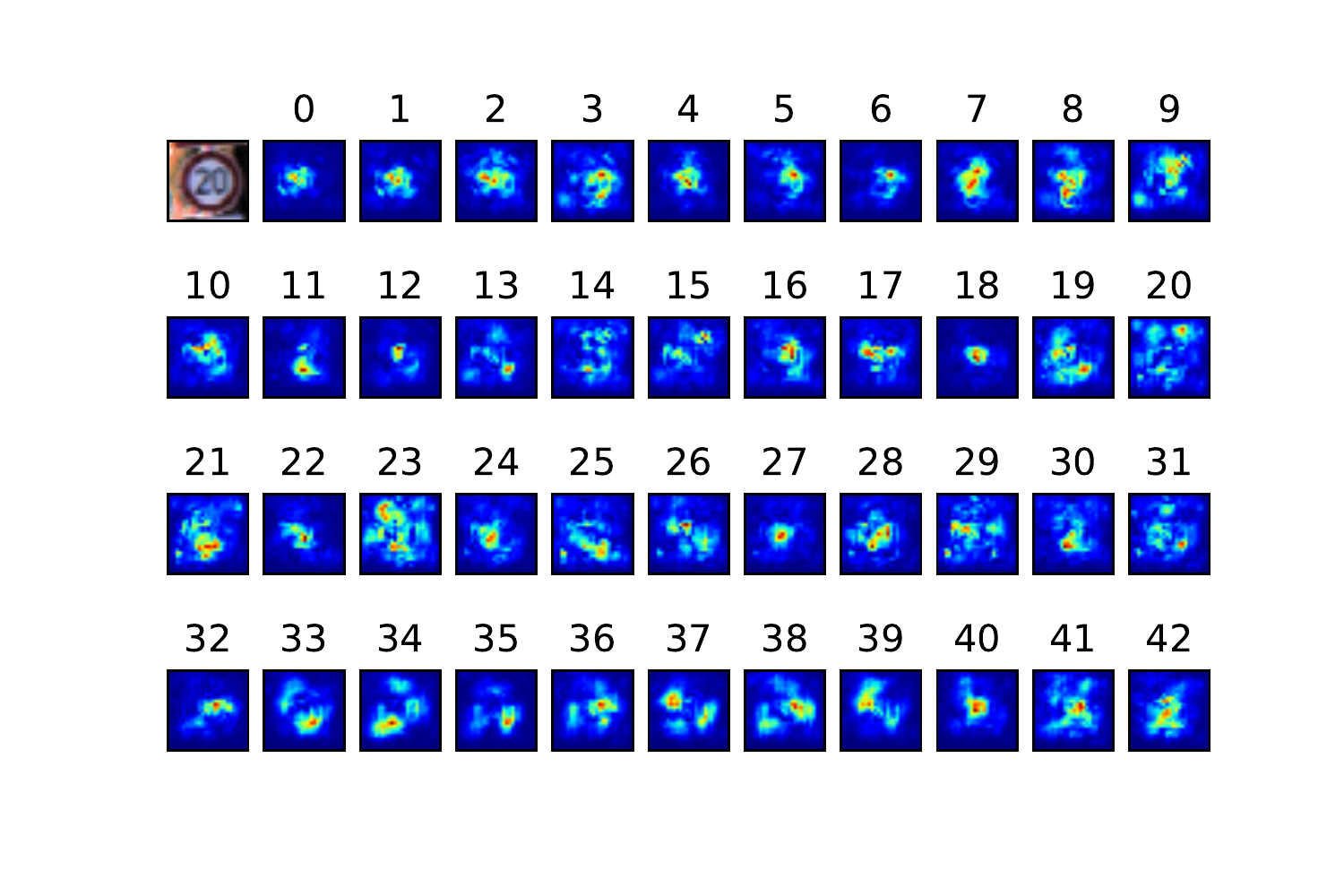}
	\end{center}
	\caption{Output explanation of image A, clean model}
	\label{Figure:clean_explaination_0}
\end{figure}

\begin{figure}[H]
	\begin{center}
		\includegraphics[width=0.35\textwidth]{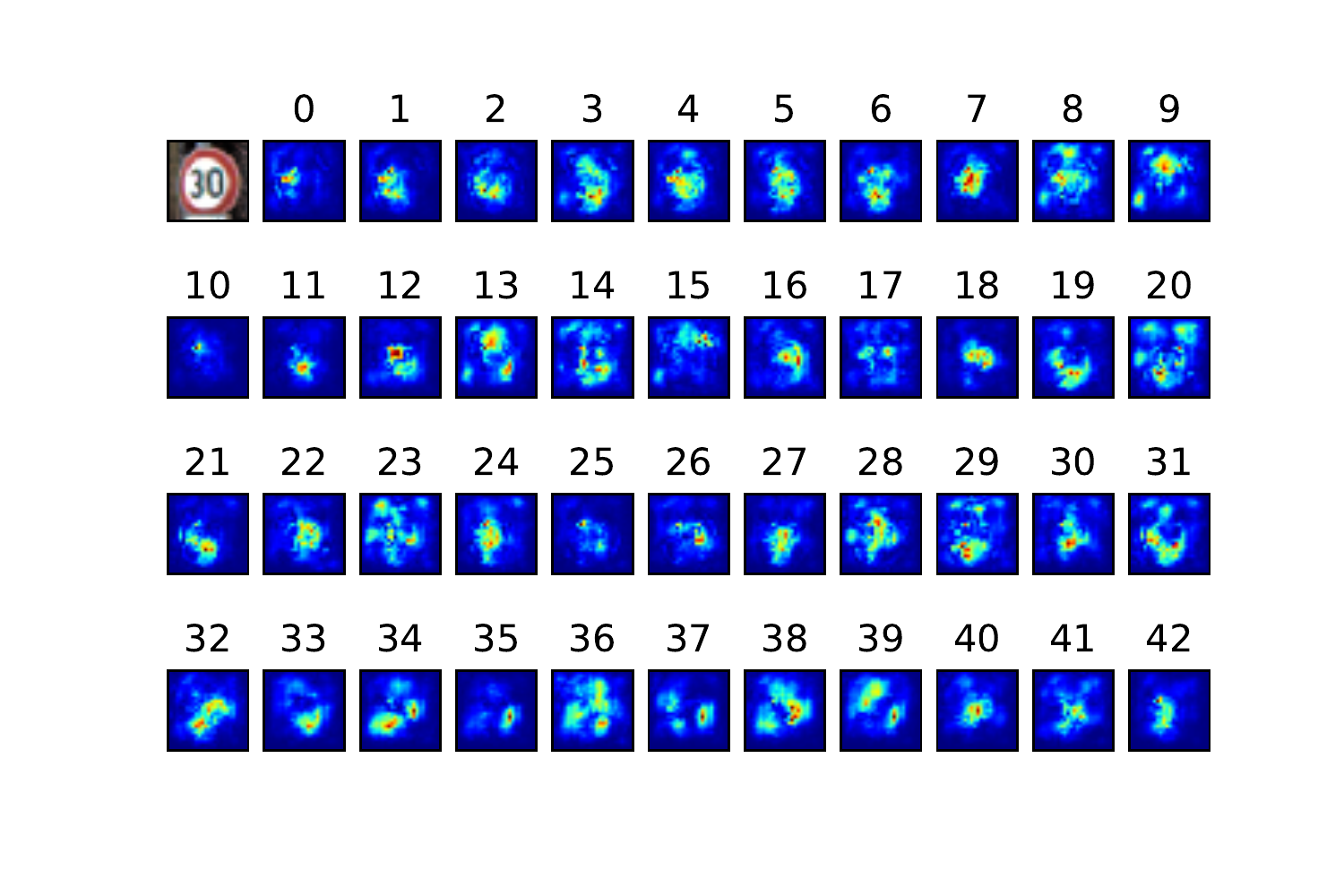}
	\end{center}
	\caption{Output explanation of image B, clean model}
	\label{Figure:clean_explaination_1}
\end{figure}

We generate the saliency heatmap with given input and a model. The saliency heatmap for different input image of a backdoor injected model is shown in Figure~\ref{Figure:backdoor_explaination_0} and Figure~\ref{Figure:backdoor_explaination_1} respectively. Notice that the attack target is class 20. From the saliency heatmap of a backdoor model, we can see that the heatmap of the attack target appears to be different from other heatmaps. The saliency heatmap for different input image of a clean model is shown in Figure~\ref{Figure:clean_explaination_0} and Figure~\ref{Figure:clean_explaination_1} respectively.
\vspace*{-5mm}
\vspace*{-5mm}
\vspace*{-5mm}
\vspace*{-5mm}

\end{document}